\documentclass[twocolumn,prb]{revtex4}
\usepackage{epsf,graphicx,amssymb,amsmath}

\begin{document}
\draft
\title{Electrical conductivity in granular media and Branly's coherer: A
simple experiment}
\author{Eric Falcon}
 \email{Eric.Falcon@ens-lyon.fr}
 \homepage{http://perso.ens-lyon.fr/eric.falcon/}
\author{Bernard Castaing}
\affiliation{Laboratoire de Physique, \'Ecole Normale Sup\'erieure de
Lyon, UMR 5672, 46, all\'ee d'Italie, 69 007 Lyon, France}

\date{\today}

\begin{abstract}
We show how a simple laboratory experiment can illustrate certain
electrical transport properties of metallic granular media. At a low
critical external voltage, a transition from an insulating to a conductive
state is observed. This transition comes from an electro-thermal coupling
in the vicinity of the microcontacts between grains where microwelding
occurs. Our apparatus allows us to obtain an implicit determination of
the microcontact temperature, which is analogous to the use of a resistive
thermometer. The experiment also illustrates an old
problem, the explanation of Branly's coherer effect -- a radio wave
detector used for the first wireless radio transmission, and based on the sensitivity of the metal fillings conductivity to an electromagnetic wave.
\end{abstract}


\maketitle

\section{Introduction}

The coherer or Branly
effect is an electrical conduction instability that appears in a
slightly oxidized metallic powder under a constraint.\cite{Branly90} The
initial high powder resistance falls irreversibly by several orders of
magnitude as soon as an electromagnetic wave is produced in its vicinity.
The effect was discovered in 1890 by E. Branly\cite{Branly90}
and is related to other phenomena. For instance, a transition from an
insulating state to a conducting state is observed as the external source
exceeds a threshold voltage (the DC Branly effect); temporal fluctuations
and slow relaxations of resistance also occur under certain
conditions.\cite{Falcon04}

Although these electrical
transport phenomena in metallic granular media were involved in the first
wireless radio transmission near 1900, they still are not well
understood. Several possible processes at the contact scale have been
invoked without a clear verification: electrical breakdown of the oxide
layers on grains,\cite{Kamarinos75} the modified tunnel effect through the
metal-oxide/semiconductor-metal junction,\cite{Holm00} the attraction of
grains by molecular or electrostatic forces,\cite{Gabillard61} and local
welding of microcontacts by a Joule heating
effect.\cite{Vandembroucq97,Dorbolo03,Holm00} A global process of percolation
within the grain assembly also has been 
invoked.\cite{Kamarinos75,Gabillard61,Vandembroucq97}

Our goal in this paper is to understand the DC Branly effect by means of an experiment with a chain of metallic beads.\cite{Falcon04bis} Our focus is on the local properties (the contacts
between grains) instead of the collective properties. We also discuss the
history of the electrical and thermal properties of non-homogeneous media
such as granular media, as well as the influence of electromagnetic waves on
their conductance.\cite{Taylor04} After a brief review of the history of the
coherer effect in Sec.~\ref{history}, we introduce in Sec.~\ref{dispo} an
experiment that can be easily done in a standard physics laboratory. We
present our results in Sec.~\ref{results}, followed by a qualitative and
quantitative interpretation of the conduction transition mechanism in
Secs.~\ref{qualitative} and
\ref{quantitative}. Our conclusions are given in Sec.~\ref{conclusion}.

\section{A brief historical review}\label{history}
In 1887, shortly after the publication of Maxwell's
theory of electromagnetism, experiments performed by H.\ Hertz clearly
demonstrated the free space generation and propagation of electromagnetic
waves. He noticed that sparks (high frequency electromagnetic waves of
the order of 100\,MHz) could induce arcing across a wire loop containing a
small gap, a few meters away.\cite{Hertz1893,Emerson01} 

This discovery was anticipated by many people; P.\ S.\ Munk observed in
1835 the permanent increase of the electrical conductivity of a mixture
of metal filings resulting from the passage of a discharge current of a
Leyden jar.\cite{Kryzhanoskii92} In 1879 D.\ E.\ Hughes observed a similar
phenomenon for a loose contact formed of a carbon rod resting in the grooves
in two carbon blocks, and with a tube filled with metallic granules (a
microphone because it was first designed to detect acoustic waves). Hughes
appears to have discovered the important fact that such a tube was sensitive
to electric sparks at a distance as indicated by its sudden change in
conductivity. At the time, the Royal Society of London was not convinced, and
his results were published some 20 years later,\cite{Hughes99} a long time
after the discovery of hertzian waves. In 1884 T.\ Calzecchi-Onesti performed
experiments on the behavior of metallic powders under the action of various
electromotive forces, and observed a considerable increase of the
powder conductivity by successively opening and closing a circuit
containing an induction coil and a tube with fillings.\cite{Fleming1906}

The action of nearby electromagnetic waves on metallic powders was
observed and extensively studied by Branly in 1890.\cite{Branly90}
When metallic filings are loosely arranged between two electrodes in a
glass or ebonite tube, it has a very high initial resistance of many
megohms due to an oxide layer likely present on the particle surfaces.
When an electric spark was generated at a distance away, the resistance was
suddenly reduced to several ohms. This conductive state remained until the
tube was tapped restoring the resistance to its earlier high value. Because
the electron was not known at this time (it was discovered in
1897),\cite{Blatt89} Branly called his device a ``radio conductor'' to
recall that ``the powder conductivity increased under the influence of the
electric radiations from the spark;'' the meaning of the prefix ``radio'' at
this time was ``radiant'' or ``radiation.'' He performed other experiments
with various powders, lightly or tightly compressed, and found that the same
effect occurred for two metallic beads in contact, and for two slightly
oxidized steel or copper wires lying across each other with light
pressure.\cite{Branly02} This loose or imperfect contact was found to be
extremely sensitive to a distant electric spark.

This discovery caused a considerable stir when O.\ Lodge in 1894
repeated and extended Hertz's experiments by using a Branly tube, a much
more sensitive detector than the wire loop used by
Hertz.\cite{Emerson01,Fleming1906} Lodge improved the Branly tube so that it
was a reliable, reproducible detector, and automated it by tapping on the
tube with a slight mechanical shock. Lodge called this electromagnetic wave
detector a ``coherer'' from the Latin {\it cohaerere}, which means ``stick
together.'' He said that the fillings ``coherered'' under the action of the
electromagnetic wave and needed to be ``decoherered'' by a shock. Later,
Branly and Lodge focused their fundamental research on mechanisms of powder
conductivity, and not on practical applications such as wireless
communications. However, based on using the coherer as a wave detector, the
first wireless telegraphy communications were transmitted in 1895 by G.\
Marconi, and independently by A.\ S.\
Popov.\cite{Bridgman01,Kryzhanoskii92,Fleming1906} Popov also used the
coherer to detect atmospheric electrical discharges at a distance.

Lodge first hypothesized that the metallic grains were welded together by
the action of the voltages that are induced by electromagnetic
waves.\cite{Fleming1906} According to some including Lodge,\cite{Tommasina98,Fleming1906} the grains became dipoles
and attracted each other by electrostatic forces, inducing  grains to stick
together, thus forming conductive chains. A shock should be enough to break
these fragile chains and to restore the resistance to its original value.
Branly did not believe this hypothesis, and to demonstrate that motion of
the grains was not necessary, he immersed the particles in wax
or resin. He also used a column of six steel balls or disks, which were a
few centimeters in diameter. Because the coherer effect persisted, he thought
that the properties of the dielectric between the grains played an important
role. In 1900, Guthe et al. performed similar experiments with two balls in
contact.\cite{Guthe01} However, the invention by de Forest of the triode in
1906, the first vacuum tube (an audion), supplanted the coherer as a
receiver, and Branly's effect sank into oblivion without being fully
understood.

In the beginning of the 1960s, a group in Lille
became interested in this old problem. They suggested that
attractive molecular forces keep the particles in contact even after the
removal of the applied electrostatic field.\cite{Gabillard61} In the 1970s,
numerous papers considered the conductivity of granular materials for
batteries, but they did not focus on the electrical
conduction transition.\cite{Euler78} In 1975, a group in Grenoble suggested a
mechanism of electrical breakdown of the oxide layer on the grain surfaces
and investigated the associated
$1/f$ resistance noise.\cite{Kamarinos75} In 1997, the conduction transition
was observed by direct visualization (with an infrared camera) of the
conduction paths when a very high voltage ($>500$\,V) was applied to a
monolayer of aluminium beads.\cite{Vandembroucq97} More recently, the action
at a distance of sparks was investigated.\cite{Dorbolo03} For additional
information about the history of the coherer, see
Refs.~\onlinecite{Histo} and \onlinecite{Contro}.

\section{Electrical conductivity of a chain of metallic beads}
Understanding the electrical conduction transition in granular
materials is a complicated problem that depends on many
parameters: the 
statistical distribution of the shape and size of the grains, the applied force, and the local properties at the contact scale of two grains,
that is, the degree of oxidization, surface state, and roughness. Among the
phenomena proposed to explain the coherer effect, it is easy to show that
some have only a secondary contribution. For instance, because the coherer
effect was observed by Branly with a single contact between two
grains,\cite{Branly02} percolation cannot be the dominant mechanism.
Moreover, when two beads in contact are connected in series with a battery,
a coherer effect is observed at a sufficiently high imposed
voltage,\cite{Guthe01} in a way similar to the action at a distance of a
spark or an electromagnetic wave. We will reduce the number of parameters,
without loss of generality, by focusing on electrical transport within a
chain of metallic beads directly connected to a DC electrical source. 

\begin{figure}[h]
\centerline{
\epsfxsize=75mm
\epsffile{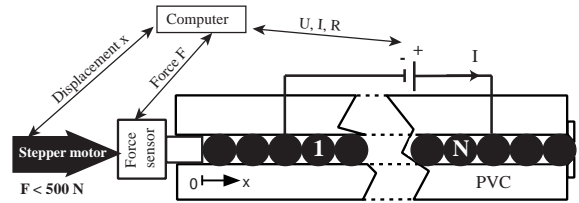}
}
\caption{Schematics of experimental setup.}
\label{fig01}
\end{figure}

\subsection{Experimental setup}\label{dispo}
The experimental setup is sketched in Fig.~\ref{fig01}. It consists of a
chain of 50 identical stainless steel beads,\cite{AISI} each 8\,mm in
diameter, and $0.1\,\mu$m in roughness. The beads are surrounded by an insulating medium of polyvinylchloride (PVC). A static force
$F\leq 500$\,N is applied to the chain of beads by means of a stepper
motor, and is measured with a static force sensor. The number of motor
steps is measured with a counter to determine $x$, the total deformation of
the chain that is necessary to reach a specific force. During a typical
experiment, we supply a current ($10^{-6}\,{\rm A} \leq I \leq 1$\,A) and
simultaneously measure the voltage $U$, and thus the resistance
$R=U/I$. Similar results have been found by repeating the experiment
with an applied voltage and measuring $I$ and thus $R$. The number of
beads 
$N$ between the two electrodes is varied from 1 to 41 by moving the
electrode beads within the chain. The
lowest resistance of the entire chain (a few ohms) is always found to be
much higher than that of the electrode and the stainless steel bulk
material.

\subsection{Experimental results}\label{results}
The mechanical behavior of the bead chain is found to be in very good
agreement with the nonlinear Hertz law (given by linear elasticity), that is
$F \propto x^{3/2}$. This result leads to an estimate of the typical range of
the deformation between two beads as 2 to
$20\,\mu$m, and of the apparent contact radius, $A$, of 40 to
$200\,\mu$m, when $F$ ranges from 10 to 500\,N.

\begin{figure}[h]
\centerline{
\epsfysize=65mm
\epsffile{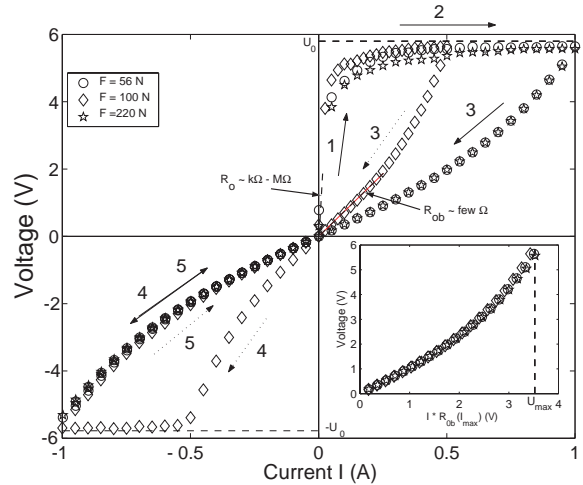}
}
\caption{Symmetrical characteristics of a chain of $N = 13$ beads for
various forces $F$ and for various current cycles in the range $-I_{\max}
\leq I \leq +I_{\max}$. ($\circ$, $\star$): $I=0\rightarrow1$A $\rightarrow
-1$A $\rightarrow0$, and ($\Diamond$) $I=0\rightarrow0.5$A $\rightarrow
-0.5$A $\rightarrow -1$A $\rightarrow 0$. A saturation voltage appears for
$U_{0}\simeq 5.8$ V corrresponding to a saturation voltage per contact
$U_{0}/(N+1)\simeq 0.4$ V. The inset shows the reversible return
trajectories rescaled by $R_{0{\rm b}}$. $U_{\max} \equiv R_{0{\rm b}} *
I_{\max}\simeq3.5$\,V. (See text for details.)}
\label{fig02}
\end{figure}

The electrical behavior is much more remarkable than the mechanical one.
Because no particular precautions were taken for the beads, an insulating film (oxide or
contaminant), a few nanometers thick, is likely present at the bead-bead
contact. When the applied current to
the chain is increased, we observe a transition from an insulating to a
conductive state as shown in Fig.~\ref{fig02}. At low applied current and
fixed force, the voltage-current
$U$-$I$ characteristic is reversible and ohmic (see arrow $1$ in
Fig.~\ref{fig02}) with a high, constant resistance, $R_0$. This resistance
($R_0 \simeq 10^4$--$10^7\,\Omega$) at low current depends in a complex way
on the applied force and on the contaminant film properties (resistivity and
thickness) at the contact location. The value of $R_0$ is determined by the
slope of the
$U$-$I$ plot at low current. As $I$ is increased further, the resistance
strongly decreases, corresponding to a bias $U_0$ independent of $I$ (see arrow $2$). As soon as this saturation voltage
$U_0$ is reached, the $U$-$I$ characteristic is irreversible if the current is decreased
(see arrow $3$). The
resistance reached at low decreasing current,
$R_{0{\rm b}}$ (the order of 1--$10\,\Omega$), depends on the 
previously applied maximum current,
$I_{\max}$. Note that the nonlinear return trajectory is reversible upon
again increasing the current, $I$, and also is symmetrical when the current applied to the chain
is reversed (see arrows 4 and 5). For different applied forces $F$ and
different values of
$I_{\max}$, we show that the return trajectories depend only on
$I_{\max}$ and follow the same reverse trajectory when $U$ is plotted
versus $I R_{0{\rm b}}$ (see the inset in Fig.~\ref{fig02}). The values of
$R_{0{\rm b}}$ are determined by the slopes of the $U$-$I$ return
trajectories at low and decreasing current (see Fig.~\ref{fig02}).

The decrease of the resistance by several orders of magnitude (from
$R_0$ to $R_{0{\rm b}}$) is similar to that of the coherer
effect with powders\cite{Branly90} and with a single
contact.\cite{Branly02,Guthe01} Note that after each cycle of the
current, the applied force is reduced to zero, and we roll the beads
along the chain axis to form new contacts for the next cycle. With this
procedure, the fall of the resistance (the coherer or Branly effect) and the
saturation voltage are always observed and are very reproducible.

The saturation voltage $U_0$ is independent of the applied force, but
depends on the number of beads between the electrodes. The saturation voltage
per contact $U_{c} \equiv U_0/(N+1)$ is found to be constant when the number of beads $N$ is varied from 1 to 41
and is on the order of 0.4\,V per contact. However, this saturation voltage
depends slightly on the bead material ($U_{c} \simeq 0.4$\,V for stainless
steel beads, $\simeq 0.2$\,V for bronze beads, and 0.3\,V for brass beads),
but is of the same order of magnitude.\cite{Falcon04bis}

\subsection{Qualitative interpretation}\label{qualitative}
\begin{figure}[h]
\centerline{
\epsfysize=65mm
\epsffile{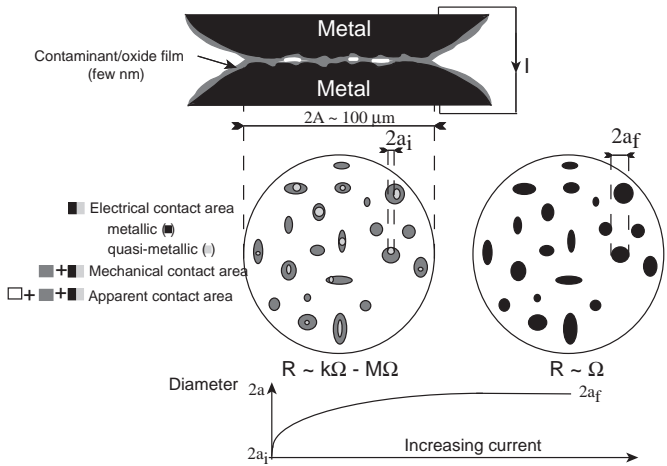}
}
\caption{Schematic of the electrical contact creation through
microcontacts by transformation of the poorly conductive
contaminant/oxide film. At low current $I$, the electrical contact is
mostly driven by a complex conduction mechanism through this film via
conductive channels (of areas increasing with $I$). At high enough $I$,
an electro-thermal coupling generates a welding of the microcontacts
leading to efficient conductive metallic bridges (of constant area).}
\label{fig03}
\end{figure}

Assume a mechanical contact between two metallic
spheres covered by a thin contaminant film ($\sim$ few nm). The interface
generally consists of a dilute set of microcontacts due to the roughness
of the bead surface.\cite{Holm00} The mean radius,
$a$, of these microcontacts is of the order of magnitude of the bead
roughness $\sim 0.1\,\mu$m, which is much smaller than the apparent
Hertz contact radius $A\sim 100\,\mu$m. Figure~\ref{fig03}
schematically shows the creation of good electrical contacts by the
transformation of this poorly conductive film. At low
applied currents, the high value of the contact resistance
(k$\Omega$--M$\Omega$) probably comes from a complex conduction path
found by the electrons through the film within the very small size ($\ll
0.1\,\mu$m) of each microcontact (see light grey zones in
Fig.~\ref{fig03}). The electrons damage the film and lead to a ``conductive
channel'': the crowding of the current lines within these microcontacts
generates a thermal gradient in their vicinity if significant Joule heat is
produced. The mean radius of the microcontacts then strongly increases by
several orders of magnitude (for example, from
$a\ll 0.1\,\mu$m to
$a\sim 10\,\mu$m), and thus enhances their conduction (see
Fig.~\ref{fig03}). This increase of the radius is responsible for the
nonlinear behavior of the $U$-$I$ characteristic (arrow $1 \to 2$ in
Fig.~\ref{fig02}). At high enough current, this electro-thermal process can
lead to local welding of the microcontacts (arrow 2 in Fig.~\ref{fig02});
the film is thus pierced in a few places where purely metallic contacts (few
$\Omega$) are created (see the black zones in Fig.~\ref{fig03}). (Note
that the current-carrying channels (bridges) are a mixture of metal and
the film material rather than a pure metal. It is likely that the
coherer action results in only one bridge -- the contact resistance is
lowered so much that piercing at other points is prevented.) The
$U$-$I$ characteristic is reversible when $I$ is decreased and then
increased (arrow 3 in Fig.~\ref{fig02}). The reason is that the
microcontacts have been welded, and therefore their final size does not
vary any more for $I < I_{\max}$. The $U$-$I$ reverse trajectory then
depends only on the temperature reached in the metallic bridge and no
longer depends on the bridge size as for the initial trajectory.

\subsection{Quantitative interpretation}\label{quantitative}
We now check the interpretation
in Sec.~\ref{qualitative} quantitatively. Assume a microcontact between
two clean metallic conductors (thermally insulated at the uniform
temperature
$T_0$, with no contaminant or tarnish film on their surfaces). Such a
clean microcontact is called a ``spot." If an electrical current
flowing through this spot is enough to produce Joule heating (assumed to
be totally dissipated by thermal conduction in the conductors), then a
steady-state temperature distribution is quickly reached ($\sim \mu$s) in
the contact vicinity. The maximum temperature reached, $T_m$, is located at
the contact, and is related to the potential, $\varphi$, at the isotherm,
$T$, by the Kohlrausch equation\cite{Holm00,Greenwood58}
\begin{equation}
\varphi^2(T)=2\!\int_T^{T_m}\lambda(T')\rho(T')dT'\ {\rm ,}
\label{Phi-T}
\end{equation}
where $\lambda(T)$ is the thermal conductivity and $\rho(T)$ is the
electrical resistivity of the conductor, both depending on the
temperature $T$. Thermal equilibrium means that the heat flux,
$\lambda(T) \vec \nabla T$, across the isothermal surfaces, $S$, is due to
the electrical power,
$I\varphi(T)$, where $\varphi(T)$ is the potential between one of the
conductors and the contact ($\varphi(T_0)=\pm U/2$). This thermal
equilibrium, $I\varphi(T)= \iint\limits_S-\lambda\vec \nabla(T).\,d\vec
S$, and the current density $\vec j=-\vec \nabla(\varphi)/\rho$, thus
gives
$\varphi d\varphi=\mp \lambda\rho\,dT$, which leads by integration to
Eq.~(\ref{Phi-T}).

For many conductors, the Wiedemann-Franz law states
that\cite{Holm00}
\begin{equation}
\lambda \rho= LT,
\label{WFlaw}
\end{equation}
where $L=\pi^2 k^2/(3e^2)=2.45\times 10^{-8}$\,V$^2$/K$^2$ is the Lorentz
constant, $k$ is the Boltzmann constant, and $e$ is the electron charge.
If we combine Eqs.~(\ref{Phi-T}) and (\ref{WFlaw}) with $\varphi(T_0)=\pm
U/2$, we can express the relation between $T_m$ and the applied voltage,
$U$, as
\begin{equation}
T_m^2 - T_0^2 = \frac{U^2}{4L}.
\label{WF}
\end{equation}
Equation~(\ref{WF}) shows that the maximum temperature
$T_m$ reached at the contact is independent of the contact geometry and
of the materials in contact because both the electrical resistivity,
$\rho(T)$, and the thermal conductivity, $\lambda(T)$, are due to the
conduction electrons, which leads to 
the temperature dependence given by Eq.~(\ref{WFlaw}).

A voltage near 0.4\,V across a contact leads, from Eq.~(\ref{WF}) and the
value of $L$, to a contact temperature near $1050^\circ$C ($T_m=1320\,$K)
for a bulk temperature $20^\circ$C ($T_0=290$\,K). A voltage $U\simeq
0.3$--0.4\,V thus leads from Eq.~(\ref{WF}) to contact temperatures that
exceed the melting point of most conducting materials. Efficient metallic
bridges are therefore created by microwelding. Beyond the quantitative
agreement with the experimental saturation voltage
$U_{c}$ (see Sec.~\ref{results} and Fig.~\ref{fig02}), Eq.~(\ref{WF})
also explains why
$U_{c}$ is the relevant parameter in the experiments in
Sec.~\ref{results}, and not the magnitude of the current. In addition,
when
$U$ approaches $U_{c}$ (see Fig.~\ref{fig02}), the local heating of the
microcontacts is enough, from Eq.~(\ref{WF}), to melt them. Then their
contact areas increase, thus leading to a decrease of the local resistance.
When $U_{c}$ is reached, the microcontacts are welded, thus stabilizing
the contact areas, the voltage, and the contact temperatures. The phenomenon
is therefore self-regulated in voltage and temperature.

Our quantitative model describes only the electrical behavior of a welded
contact, that is, when the saturation voltage is reached. It describes the reversible $U$-$I$ reverse trajectory (when this
contact is cooled by decreasing the current from $I_{\max}$, then eventually
reheated by increasing $I$.) The contact area is assumed to be constant
because the contact has been welded, and $I < I_{\max}$.

Let us derive the analytical expression of the nonlinear $U$-$I$ reverse
trajectory.\cite{Falcon04bis} We introduce the ``cold'' contact resistance
$R_{0{\rm b}}$ present at currents sufficiently low so as to not cause any
appreciable rise in the temperature at the contact. The bulk conductor 
is at the room temperature $T_0$, with an electrical resistivity
$\rho_0 =
\rho(T_0)$. The derivation of $R_{0{\rm b}}$ involves the same
equipotential surfaces during a change between the ``cold'' state
(denoted by a star) $\varphi^{\star}$, and the ``hot'' state
$\varphi(T)$: the same current thus involves the same current density in
both states, and thus $\vec \nabla (\varphi^{\star})/\rho_0=\vec
\nabla(\varphi)/\rho(T)$. Note that an equipotential also is an isothermal.
At thermal equilibrium, this equation and the differential expression of
Eq.~(\ref{Phi-T}) give 
\begin{equation}
\frac{d\varphi^{\star}}{\rho_0}=\frac{d\varphi}{\rho(T)}=
\mp \frac{\lambda(T)}{\varphi(T)}dT .
\label{Equipot}
\end{equation}
We use Ohm's law and integrate Eq.~(\ref{Equipot})
between the isothermal surfaces $T_0$ and $T_m$ and find\cite{Greenwood58}
\begin{equation}
\frac{IR_{0{\rm
b}}}{\rho_0}=2\!\int_{T_0}^{T_m}\frac{\lambda(T)}{\varphi(T)}dT.
\label{Rob}
\end{equation}
The factor of two arises from heat flowing in parallel on both sides of the contact whereas the current uses these both sides in series. The temperature dependence of the
thermal conductivity,
$\lambda(T)$, and electrical conductivity, $\rho(T)$, of the material in
contact is given by Eq.~(\ref{WFlaw}) and
\begin{equation}
\rho(T) = \rho_0[1+\alpha(T - T_0)],
\label{rho}
\end{equation}
where $\alpha$ is the temperature coefficient of the electrical
resistivity.

Equations~(\ref{WFlaw}) and (\ref{rho}) let us find explicit
expressions for 
$\lambda(T)$ and $\varphi(T)$ which can be substituted in Eq.~(\ref{Rob}).
The reverse trajectory $IR_{0{\rm b}}$ depends only on the temperature
$T_m$ (that is, on $U$),
\begin{equation}
T_m=\sqrt{T_0^2+\frac{U^2}{4L(N+1)^2}}, 
\label{solT}
\end{equation}
and finally gives (see the Appendix in Ref.~\onlinecite{Falcon04bis} for
the details)
\begin{equation}
IR_{0{\rm
b}}=2(N+1)\frac{\sqrt{L}}{\alpha}\!\int_0^{\theta_0}
\frac{\cos{\theta}}{[1+(\alpha T_0)^{-1}]
\cos{\theta_0}+\cos{\theta}}d\theta,
\label{solIalloy}
\end{equation}
where $\theta_0 \equiv \arccos{(T_0/T_m)}$ and $N+1$ is the number of
contacts in series in the chain. Note that only $R_{0{\rm b}}$ depends on
the contact geometry, and its value is easily determined experimentally (see
Sec.~\ref{results}).

\begin{figure}[ht]
\centerline{
\epsfysize=65mm
\epsffile{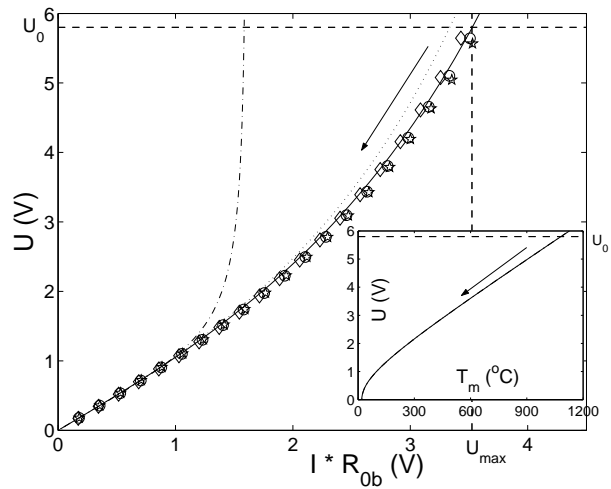}
}
\caption{Comparison between the experimental $U$-$I$ reverse trajectories
of Fig.~\ref{fig02} (symbols) and theoretical curves from
Eq.~(\ref{solIalloy}) for an alloy with stainless steel
properties [$\alpha^{-1}=4T_0$ ($-$) or 
$3.46T_0$ ($\cdots$)], and for a pure metal [$\alpha^{-1}=T_0$ ($-.-$)].
The inset shows the theoretical maximum temperature, $T_m$, from
Eq.~(\ref{solT}), reached for one contact when the chain of $N=13$
stainless steel beads is submitted to a voltage
$U$.}
\label{fig04}
\end{figure}

Because for pure metals ($\alpha^{-1}\simeq T_0$),\cite{Handbook} the
right-hand side of Eq.~(\ref{solIalloy}) does not depend explicitly on the
geometry of the contact or on the metal used for the contact. However, for
alloys the right-hand side of Eq.~(\ref{solIalloy}) depends on $\alpha$, the
temperature coefficient of the electrical resistivity of the alloy. This
additional parameter is related to the presence of defects in the
material. The normalized
$U$-$I$ reverse trajectory (that is, $U$ as a function of
$IR_{0{\rm b}}$ in the inset of Fig.~\ref{fig02}) are compared in
Fig.~\ref{fig04} with the theoretical solutions, Eq.~(\ref{solIalloy}),
for pure metals and for a stainless steel alloy. Very good
agreement is found between the experimental results and the
electro-thermal theory, especially for the alloy. Qualitatively, the
alloy solution is closer to the experimental data than the solution
for a pure metal. The agreement is even quantitatively excellent (see the
solid line in Fig.~\ref{fig04}). For this comparison, the value 
$\alpha^{-1}$ is equal to $4T_0$ instead of $3.46T_0$ (the
$\alpha^{-1}$ value for AISI 304 stainless steel),\cite{Steel} because the
value of $\alpha^{-1}$ for the bead material (AISI 420 stainless steel)
is unknown, but should be close to $3.46T_0$. During the
experimental reverse trajectory, the equilibrium temperature, $T_m$, of a
microcontact also is deduced from Eq.~(\ref{solT}) with no adjustable
parameters (see the inset in Fig.~\ref{fig04}). Therefore, when the
saturation voltage is reached ($U_0=5.8$\,V), $T_m$ is close to
$1050^\circ$C which is enough to soften or melt the microcontacts between
the $N=13$ beads of the chain. We could say that our implicit measurement
of the maximum temperature (based on the temperature dependence of the
material conductivities) is equivalent to the use of a resistive thermometer.

\section{Conclusions}\label{conclusion}

Electrical phenomena in granular materials related to the electrical conduction
transition such as the Branly effect have been interpreted in many
different ways but without a clear demonstration. We have reported the
observation of electrical transport through a chain of oxidized metallic
beads under an applied static force. A transition from an insulating to a
conducting state is observed as the applied current is increased. The
$U$-$I$ characteristics are nonlinear, hysteretic, and saturate to a low
voltage per contact ($\simeq 0.4$\,V). From this simple
experiment, we have shown that the transition triggered by the
saturation voltage arises from an electro-thermal coupling in the
vicinity of the microcontacts between each bead. The current flowing
through these spots generates local heating which leads to an increase of
their contact areas, and thus enhances their conduction. This
current-induced temperature rise (up to $1050^\circ$C) results in the
microwelding of contacts (even for a voltage as low as 0.4\,V). Based on
this self-regulated temperature mechanism, an analytical expression for
the nonlinear $U$-$I$ reverse trajectory was derived, and was found to be
in good agreement with the data. The theory also
allows for the determination of the microcontact temperature through
the reverse trajectory with no adjustable parameters. We could attempt to
directly visualize this process with a
microscope or infrared camera. But for this purpose a very powerful
electrical source must be applied, far in excess of that necessary to
produce true coherer phenomena (see for example
Ref.~\onlinecite{Vandembroucq97}).

\begin{acknowledgements}
We thank D.\ Bouraya for the realization of the experimental setup, and
Madame M.\ Tournon-Branly, the granddaughter of E.\
Branly, for discussions.
\end{acknowledgements}

\end{document}